\documentclass[10pt,a4paper,twocolumn,english,aps,prl,twocolumn,showpacs,floatfix,superscriptaddress,10pt]{revtex4-1}
\usepackage[T1]{fontenc}
\usepackage[latin9]{inputenc}
\usepackage{color}
\usepackage{babel}
\usepackage{amstext}
\usepackage{amssymb}
\usepackage{graphicx}
\usepackage[unicode=true,
 bookmarks=true,bookmarksnumbered=false,bookmarksopen=false,
 breaklinks=false,pdfborder={0 0 0},backref=false,colorlinks=true]
 {hyperref}
\hypersetup{pdftitle={Perpendicular  magnetization reversal in Pt/[Co/Ni]3/Al multilayers \textit{via} the Spin Hall Effect of Pt},
 pdfauthor={Carlos ROJAS}}

\makeatletter

\@ifundefined{textcolor}{}
{%
 \definecolor{BLACK}{gray}{0}
 \definecolor{WHITE}{gray}{1}
 \definecolor{RED}{rgb}{1,0,0}
 \definecolor{GREEN}{rgb}{0,1,0}
 \definecolor{BLUE}{rgb}{0,0,1}
 \definecolor{CYAN}{cmyk}{1,0,0,0}
 \definecolor{MAGENTA}{cmyk}{0,1,0,0}
 \definecolor{YELLOW}{cmyk}{0,0,1,0}
}

\@ifundefined{date}{}{\date{}}
\makeatother

\begin{document}

\title{Perpendicular  magnetization reversal in Pt/[Co/Ni]$_3$/Al multilayers \textit{via} the Spin Hall Effect of Pt}

\author{J.-C. Rojas-Sánchez}
\email[]{juan-carlos.rojas-sanchez@univ.lorraine.fr}
\affiliation{Unité Mixte de Physique, CNRS, Thales, Univ. Paris-Sud, Université Paris-Saclay, 91767 Palaiseau, France }

\author{P. Laczkowski}
\affiliation{Unité Mixte de Physique, CNRS, Thales, Univ. Paris-Sud, Université Paris-Saclay, 91767 Palaiseau, France }

\author{J. Sampaio}
\affiliation{Laboratoire de Physique des Solides, Univ. Paris-Sud, Université Paris-Saclay, 91405 Orsay, France }

\author{S. Collin}
\affiliation{Unité Mixte de Physique, CNRS, Thales, Univ. Paris-Sud, Université Paris-Saclay, 91767 Palaiseau, France }

\author{K. Bouzehouane}
\affiliation{Unité Mixte de Physique, CNRS, Thales, Univ. Paris-Sud, Université Paris-Saclay, 91767 Palaiseau, France }

\author{N. Reyren }
\affiliation{Unité Mixte de Physique, CNRS, Thales, Univ. Paris-Sud, Université Paris-Saclay, 91767 Palaiseau, France }

\author{H. Jaffrès }
\affiliation{Unité Mixte de Physique, CNRS, Thales, Univ. Paris-Sud, Université Paris-Saclay, 91767 Palaiseau, France }

\author{A. Mougin}
\affiliation{Laboratoire de Physique des Solides, Univ. Paris-Sud, Université Paris-Saclay, 91405 Orsay, France }

\author{J.-M. George}
\email[]{jean-marie.george@thalesgroup.com}
\affiliation{Unité Mixte de Physique, CNRS, Thales, Univ. Paris-Sud, Université Paris-Saclay, 91767 Palaiseau, France }

\date{\today}
\begin{abstract}
We experimentally investigate the current-induced magnetization reversal in Pt/[Co/Ni]$_3$/Al multilayers combining the anomalous Hall effect and magneto-optical Kerr effect techniques in crossbar geometry. The magnetization reversal occurs through nucleation and propagation of a domain of opposite polarity for a current density of the order of 0.3 TA/m$^2$. In these experiments we demonstrate a full control of each stage: i)the {\O}rsted field controls the domain nucleation and ii) domain-wall propagation occurs by spin torque from the Pt spin Hall effect. This scenario requires an in-plane magnetic field to tune the domain wall center orientation along the current for efficient domain wall propagation. Indeed, as nucleated, domain walls are chiral and Néel like due to the interfacial Dzyaloshinskii-Moriya interaction.
\end{abstract}
\maketitle

Controlling the magnetization reversal using the spin transfer torque (STT) is a key ingredient for the implementation of several technologies, including for example MRAM, benefiting from the scalability of the effect \cite{Boulle2011}. However, in materials stacks using conventional STT, the magnetization reversal requires a high current density of the order of 10$^{10}$ A/m$^2$, which should ideally be as small as possible in order to avoid detrimental aftereffects when flowing across tunnel junctions based-MRAM \cite{Prejbeanu2013}. An alternative route for magnetization control is by means of the so-called spin-orbit torque (SOT), which uses materials with a strong spin-orbit coupling (SOC) such as Pt, to generate spin currents along the perpendicular directions to the charge current \textit{via} its spin Hall effect (SHE). It is then possible to take advantage of this spin current with the only difference that now both charge and spin currents are orthogonal to each other in a typical multilayer system. Charge current is flowing along the interface while spin current is diffusing perpendicularly to the interface of the constitutive layers.

In this vein, several experiments have been designed to demonstrate the possibility of reversing the magnetization in single ferromagnetic layer or to propagate domain walls (DWs) using materials with strong SHE \cite{Miron2011,Liu2012a,Ryu2012,Kim2012,Emori2013a,Ryu2013a}. Unfortunately the exact origin of the process as well as its microscopic understanding were still far from being fully understood at the present stage\cite{Thiaville2012,Khvalkovskiy2013,Martinez2014a,Perez2014,Kundu2015}. Since the pioneering work from Miron \textit{et al}., \cite{Miron2011} several reports have shown some possible routes or scenarios to explain either magnetization reversal \cite{Liu2012a,Bi2014,Lee2014a,Qiu2014,Hao2015a,Bi2015a,Bhowmik2015,Zhang2015b,Yang} or DW propagation \cite{Ryu2012,Kim2012,Emori2013a,Ryu2013a,Torrejon2014b,Pizzini2014,Emori2014,Vanatka2015,LoConte2015,SanEmeterioAlvarez2010,Chen2013}. The latest involve DW propagation taking into account the Dzyaloshinski-Moriya interaction (DMI) at non magnetic/magnetic interfaces in magnetic multi-domain configurations \cite{Thiaville2012,Emori2013a,Ryu2013a,Lee2014a,Qiu2014,Hao2015a,Bi2015a,Bhowmik2015,Yang,Torrejon2014b,Pizzini2014,Emori2014,Vanatka2015,LoConte2015}. One common point of these studies is the use of magnetic materials with perpendicular magnetic anisotropy (PMA) as well as the application of a small in-plane magnetic field along the current direction to electrically reverse the magnetization \cite{Miron2011,Liu2012a,Bi2014,Lee2014a,Qiu2014,Hao2015a,Bi2015a,Bhowmik2015,Zhang2015b,Yang}. In most of these experiments, the critical current to switch the perpendicular magnetization was supposed to be proportional to the spin Hall angle of the heavy nonmagnetic (NM) layer.
 The produced spin current exerts a torque on the magnetization when it is absorbed by the ferromagnetic layer in contact.

In this letter, we present a series of measurements of both, anomalous Hall effect (AHE) and magneto-optical Kerr effect (MOKE microscopy) in $//$Pt$/$[Co$/$Ni]$_3/$Al multilayers patterned in Hall crossbar(grown on oxidized Si-SiO$_2$ substrate). We first show that the nucleation process occurs at the edge of the wires where is applied the charge current due to the {\O}rsted field. This demonstrate that the critical switching current also depends on the {\O}rsted field. The study of DW propagation support the existence of a Néel DW configuration at zero field, due to the DMI spin-orbit interaction at the Co$/$Pt interface. A proper in-plane magnetic field is needed to reorient the centre of the DW, allowing it to efficiently propagate and fully reverse the magnetization.

The system under study consists of $//$Pt(6)$/$[Co(0.2)$/$Ni(0.6)]$_3/$Al(5) multilayers grown by dc magnetron sputtering. The numbers in parenthesis stand for thicknesses in nm. We intentionally chose a capping layer of 5 nm of Al (which will naturally oxidize over typically 2 or 3 nm) to rule out any supplementary contribution related to the oxidation at Ni$/$Al interfaces in order to focus our discussion on Pt$/$Co and Co$/$Ni interfaces. It has been shown that Co$/$Ni exhibits a perpendicular magnetic anisotropy whose origin lies in the hybridization and spin-polarized charge transfer at Co$/$Ni interfaces \cite{Posth2009,Beaujour2009a,Fukami2010,Mizukami2011,Zhang2014b}. Similar systems that have been reported to promote PMA are Co$/$Pt$/$Ni multilayers \cite{Tanigawa2012} and Pt$/$Co$/$Pt \cite{Haazen2013} trilayers. Our samples consist of 3 periods of Co$/$Ni bilayers on top of Pt which increase further the PMA. Under reasonable growth conditions, low roughness and a well defined (1 1 1) crystalline texture can be obtained in Pt, which is mandatory to induce a large PMA in Co$/$Ni systems. Pt is a well-established SHE material with a low resistivity. We have previously characterized the properties of our sputtered Pt by inverse SHE and spin pumping - ferromagnetic resonance in Co$/$Pt and Co$/$Cu$/$Pt structures for which we found a spin Hall angle, $\theta_{\textrm{SHE}} = 0.056$ for a resistivity of $\rho= 17.3$ $\mu\Omega$·cm (spin Hall conductivity of 3.2 kS/cm), and a spin diffusion length, $l_{\textrm{sf}} = 3.4$ nm \cite{Rojas-Sanchez2014}. In the present case, we varied both the [Co$/$Ni] repetition number and the Pt thickness in the $//$Pt$/$[Co$/$Ni]$_{N}/$Al multilayers in order to measure the sheet resistance and estimate the resistivity of the Co$/$Ni bilayers and the Pt layer. It results in a resistivity of 32.1 $\mu\Omega$·cm for Co$/$Ni, similar to the one reported in ref. \cite{Zhang2014b}, and 20 $\mu\Omega$·cm for Pt, close to previously reported values \cite{Rojas-Sanchez2014,Cormier2010}. The magnetization at saturation, measured using a SQUID magnetometer, is $Ms=540$ kA/m. 
The samples were patterned using UV lithography in shapes of Hall crossbars  with widths varying between 4 and 20 $\mu$m. The PMA constant $K_1 = 0.34$ MJ/m$^3$ was extracted by performing an out-of-plane angular dependence of the AHE resistance $R_\textrm{AHE}$ and the numerical fit to the experimental data.

\begin{figure}
\begin{centering}
\includegraphics[width=8.0cm]{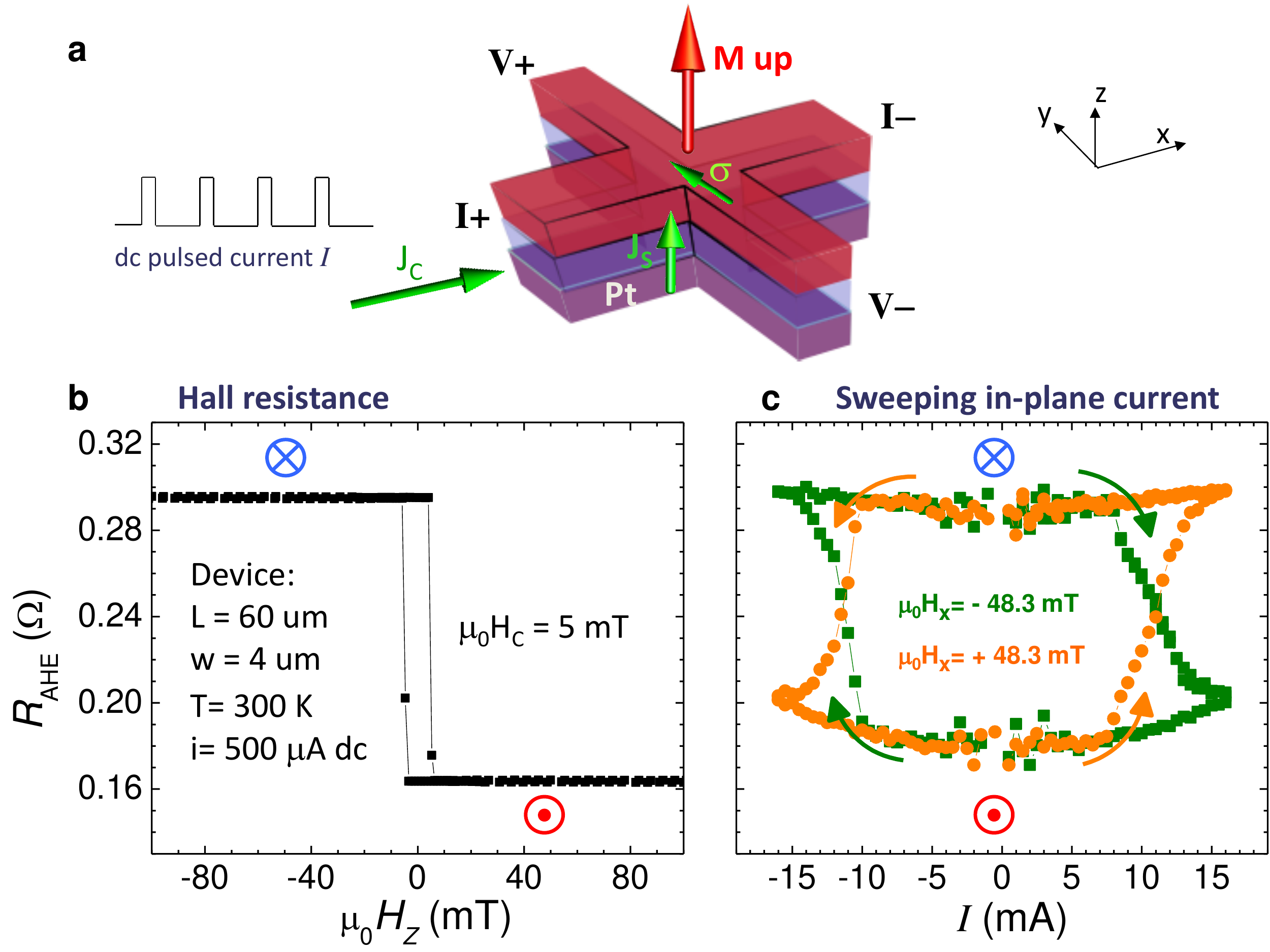}
\par\end{centering}

\caption{\label{Fig1_AHE}(color online) (a) Scheme of the Hall cross patterned in $//$Pt$/$[Co/Ni]$_3/$Al multilayer. The labelled electrical connections correspond to the AHE measurement. (b) AHE resistance measured in a device with 4 $\mu$m width at room temperature when sweeping magnetic field perpendicular to the film plane. (c) $R_\textrm{AHE}$ measured sweeping amplitude of in-plane pulsed charge current  (200 $\mu$s) with a fixed in-plane magnetic field along x of 48.3 mT or -48.3 mT. The blue (red) vector pointing out (in) the image plane stands for magnetization pointing up (down) as shown in (a).}
\end{figure}

We discuss now the first part of the study consisting of magnetization reversal probed by AHE measurements. The measurement configuration is presented in Fig.\ref{Fig1_AHE}a. The shape of the $R_\textrm{AHE}(H_z)$ loop in Fig.\ref{Fig1_AHE}b is an evidence of the PMA. This measurement is performed using a dc current density small enough in order to let the magnetization unaffected (about 10$^9$ A/m$^2$). In order to limit Joule heating effects during the AHE measurement as a function of charge current, a dc pulse current of typically 200$\mu$s was used (Keithley 2162 source meter coupled to a Keithley 2182 nanovolmeter). As shown in Fig.\ref{Fig1_AHE}c, this hysteresis loop is reproduced with the same amplitude of $R_\textrm{AHE}$ as in Fig.\ref{Fig1_AHE}b. In this case an in-plane charge current is swept instead of a perpendicular magnetic field, evidencing that the magnetization reversal is possible electrically. However an external in-plane magnetic field $H_x$ of 48.3 mT parallel to the charge current is needed to allow the reversal probed by AHE. Note that when the sign of the external $H_x$ is reversed, the switching polarity is reversed too (Fig.\ref{Fig1_AHE}c).

\begin{figure}
\begin{centering}
\includegraphics[width=8.0cm]{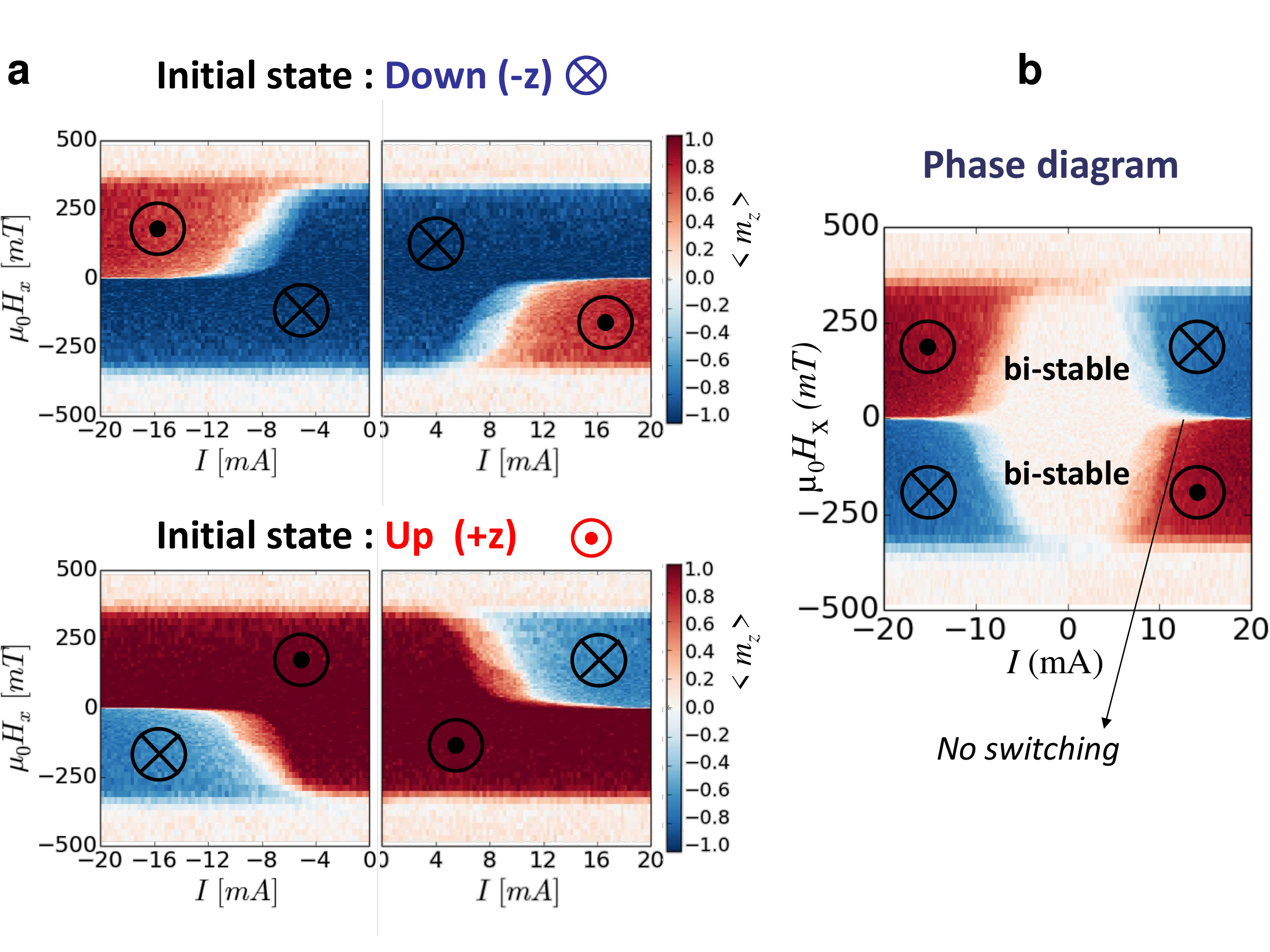}
\par\end{centering}

\caption{\label{fig:AHE_PhaseDiag}(color online) (a) Maps (color codes the average out-of-plane magnetization) of the switching conditions in a $//$Pt$/$[Co$/$Ni]$_3/$Al Hall crossbar for a given initial state (up or down magnetization) and a fixed $\mu_0H_x$ when sweeping the in-plane charge current from zero. Note that the magnetization reversal happens only under certain conditions. (b) Experimental phase diagram of the current-induced magnetization switching obtained from the four 2D maps in (a).  The central part of this image represents a bi-stable or hysteretic zone. For very low in-plane field, $\mu_0$H$_x$ $\leq$ 2.5 mT, it is not possible to induce a complete reversal of $m_z$.} 
\end{figure}

The experimental magnetization switching conditions can be further studied in detail using the following protocol: an initial state, either up or down, is prepared by applying a perpendicular magnetic field $|H_z| > |H_c|$. Then the $H_z$ is removed and a fixed value of $H_x$ is applied. The in-plane dc pulsed current from 0 to $I_\textrm{max}$ ($I_\textrm{min}$) is then swept while keeping $H_x$ constant. Fig.\ref{fig:AHE_PhaseDiag}a displays the two-dimensional color plots of the four cases, starting with an up (down) configuration and increasing (decreasing) the current to $I_\textrm{max}$ ($I_\textrm{min}$). By combining the four different experiments the experimental phase diagram of the switching conditions can be drawn (Fig.\ref{fig:AHE_PhaseDiag}b). Considering that 3 nm of Al is oxidized, the 2 nm remaining metallic Al exhibits a high resistivity (above 70 $\mu \Omega$·cm) compared to Pt (20 $\mu \Omega$·cm) and Co$/$Ni (32 $\mu \Omega$·cm), we hence neglect the current flow in Al. Therefore, for the total charge current of 10 mA flowing in the Hall crossbar of 4.5 $\mu$m width presented here, the current density in Pt and Co$/$Ni is calculated to be 0.3 TA/m$^2$ and 0.2 TA/m$^2$, respectively. This corresponds roughly to the current density value which is needed in order to reverse the magnetization. The critical current density in our system is slightly lower than the reported values for DW propagation by standard STT in FM layers \cite{Boulle2011,KoyamaT.2012}. It therefore might suggest that the origin of the magnetization reversal is controlled by the SOT coming from the SHE of Pt as evidenced hereafter.

Besides this, one can also observe that the current-induced full reversal of the magnetization happens in the range of $2.5 ~\textrm{mT}\leq\mu_0{H}_x\leq 370 ~ \textrm{mT}$ close to the saturation field at which the magnetization lies in the plane or the $\langle m_z\rangle=0$. It is easy to determine $\mu_0H_x^{max}$ experimentally performing $R_\textrm{AHE}(H_x)$ after saturating the Hall crossbar either along $(+z)$ or $(-z)$. Below the minimum $\mu_0H_x$ of 2.5 mT the system seems to reach a multidomain configuration and does not switch completely for any applied current in the limited range of $j<6\times10^{11}$A/m$^2$ to preserve the sample from damage.

We have also performed a similar series of experiments applying the external magnetic field in-plane but perpendicular to the pulsed current, $H_y$, finding that it is not possible to reverse the magnetization. Only an intermediate state (multidomain configuration) is stabilized.

To gain insight into the exact reversal mechanism process, we performed MOKE microscopy on similar devices (10 $\mu$m) exhibiting the same electrical characterizations. Same conditions have been used for MOKE experiments: four current pulses of 200 $\mu$s in duration at a given external $H_x$ and then varying the current pulse amplitude.
Examples of the acquired MOKE images after positive saturation (black color) are given after applying successive positive current pulses with no external in-plane field (Fig.3a) and two increasing in-plane field (Fig.3b and c). We observe a multidomain configuration, which rules out the occurrence of a macrospin-like reversal model (i.e. a uniform reversal of the magnetization). On the contrary, under current the reversed domains (-z, brighter colors) nucleate mostly along the bottom edge of the stripe, and propagate longitudinally (along the current), never fully reversing the stripe. If either the initial magnetization state (along z) or the direction of the current is reversed, we observe that the nucleation occurs on the top edge instead. This leads us to consider the role of the {\O}rsted field. For a rightwards current, as is the case of Fig.3a, the {\O}rsted field is pointing down (-z) at the bottom edge and +z at the top edge. In all observations, the nucleation always occurs at the edge where the {\O}rsted field is anti-parallel to the initial magnetization. We estimate that it reaches about 2 mT at the top and bottom borders for a current of 22 mA, which is close to the measured coercivity (5 mT), in Fig.\ref{Fig1_AHE}b. This strongly suggests that the nucleation location is determined by the {\O}rsted field generated by the in-plane charge current flowing along the stripe.

\begin{figure*}
\begin{centering}
\includegraphics[width=13cm]{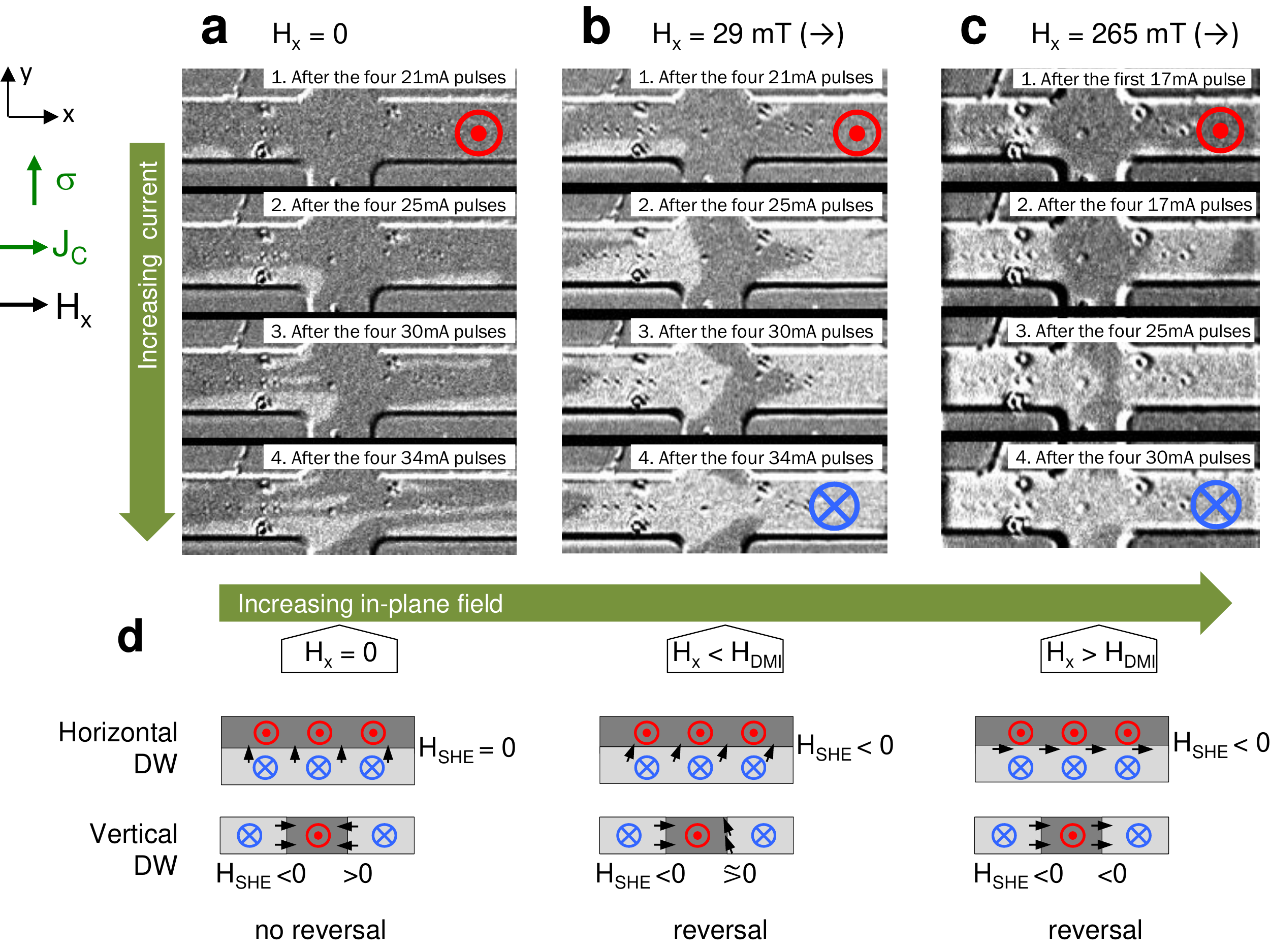}
\par\end{centering}

\caption{\label{fig:MOKE}MOKE images and schemes of the mechanism of the current-induced magnetization reversal in $//$Pt$/$[Co$/$Ni]$_3/$Al. The width of the track is 10 $\mu$ m. a-c: Kerr images after application of current pulses of increasing amplitude, four 200-$\mu$s-pulses for each value of current. The initial magnetization is +z and the charge current flows rightwards. d: Schematic of the DW and their local magnetization $m$ (small black arrows), entering $H_{SHE}\propto \sigma \times m$. Reversal occurs by propagation if $H_{SHE}<0$. Horizontal walls: reversal as soon as tilted $m$. For intermediate state in verticals DW (H<H$_{DMI}$), the reversal occurs due to the different velocities propagation of chiral N\'{e}el DW. See the text for more details and SM for the corresponding video.}
\end{figure*}

In Fig.\ref{fig:MOKE}a we also observe that the injected DWs move rightwards which is the direction of the applied current $J_\textrm{C}$ (and thus against the electron flow), ruling out the sole standard STT, as it would drive the DW in the opposite direction \cite{Emori2013a,Ryu2013a,Torrejon2014b}. Instead, the fact that the DW, once nucleated, moves only longitudinally is compatible with the SOT generated by the SHE spin currents from the Pt layer and injected into the Néel DW of a fixed chirality. The SHE-induced SOT can be written as an equivalent field $H_{\textrm{SHE}}\propto \sigma \times m$ , where $\sigma$ is the spin polarization parallel to $-$y (resulting from a spin current along +z) (see Fig.\ref{Fig1_AHE}a), and $m$ is the local magnetic moment. Néel DW have their magnetization perpendicular to the DW plane: in a \textit{vertical} Néel DW, $m$ is parallel to x, while in an \textit{horizontal} Néel DW $m$ is parallel to y (see drawings of Fig.\ref{fig:MOKE}d). Thus without any external in-plane magnetic field, $H_\textrm{SHE}$ will be non-zero and along the z direction only in a vertical DW geometry. As a consequence only vertical DW will propagate under current injection and $H_x=0$, as observed in our experiments. As Néel DW imply higher demagnetization energies than Bloch DW (in the used geometries), their presence reveals a significant interfacial DMI, which favors chiral Néel DW. The magnitude of DMI required to stabilize Néel DW compare to Bloch DW is\cite{Thiaville2012}:

\begin{equation}
D_{\textrm{crit}}=\frac{2}{\pi }{\mu_0}{H_{\textrm{NB}}}{M_{\textrm{s}}}\Delta \ ,
\end{equation}

\noindent where $\mu_0H_\textrm{NB}$ is the field needed to turn the DW configuration from Bloch to Néel in the absence of DMI \cite{KoyamaT.2012}, $M_s$ is the saturation magnetization and $\Delta$ is the DW thickness parameter $\Delta=\sqrt{A/K_\textrm{eff}}$ . Knowing that the exchange stiffnes constant is about $A = 15$ pJ/m, $K_\textrm{eff}=K_1-\mu_0M_s^2/2=0.16$  MJ/m$^3$, $\Delta=9.8$ nm, and $\mu_0H_\textrm{NB}=30$  mT (determined from micromagnetic simulations) then the minimum DMI to stabilize Néel DW configurations in our system is $D_{\textrm{crit}}=0.10$  mJ/m$^2$. Using the recent experimental DMI determination at Co/Pt interface by Brillouin Light Scattering (BLS) \cite{Belmeguenai2015} normalized to the overall total 2.4 nm FM thickness in our system, we have estimated a value of $D \approx 0.7$ mJ/m$^2$, which is much larger than $D_\textrm{crit}$. This value is in between those recent reported values in Co-Ni based system with planar (0.44 mJ/m$^2$) \cite{Di2015} or perpendicular magnetic anisotropy (0.8 mJ/m$^2$) \cite{Boulle2013a,Ryu2013a,Ryu2012}. 
This justifies the scenario illustrated in Fig.\ref{fig:MOKE}, and the reversal mechanism that we explained above and which also agrees with the simulations performed by Khvalkovskiy \textit{et al}. \cite{Khvalkovskiy2013}.

\textit{What is the role of the in-plane magnetic field}, $H_x$?
 Fig.\ref{fig:MOKE}b shows the MOKE images after positive current pulses of increasing amplitude with a positive (+x) $H_x = 29$ mT and an initial positive (+z) saturation. As with the previous case of $H = 0$, nucleation occurs at the bottom edge due to the {\O}rsted field. However, now the horizontal DW propagates and the stripe reverses. This can be understood by considering that the applied field $H_x$ rotates the magnetic moment of the horizontal DW away from the (perfect) Néel configuration (see second drawing of Fig.\ref{fig:MOKE}d). The rotation angle will be determined by the balance between $H_x$ and the DMI \cite{Boulle2013a}, which can be written as an equivalent field $\mu_0H_\textrm{DMI} = D/(M_s\Delta)$ and for our system would be $\mu_0H_\textrm{DMI} = 133$ mT. The equivalent $H_\textrm{SHE}$ on these DW is then along $-$z, causing the propagation of the horizontal DW in the direction that favors the expansion of the reversed domain. The vertical DW still propagates rightwards. However, now the in-plane field changes their velocities \cite{Hrabec2014}, which results also in a net expansion of the reversed domain. Finally, the case for large $H_x > H_\textrm{DMI}$ is shown in Fig.\ref{fig:MOKE}c (and the third drawing of Fig.\ref{fig:MOKE}d): The in-plane magnetic field overcomes $H_\textrm{DMI}$, and all DW have their magnetic moment aligned along +x, and thus all DW contribute to the expansion of the reversed domain.

In summary, we have shown the possibility of the current-induced magnetization switching in the $//$Pt$/$[Co$/$Ni]$_3/$Al system exploiting the SHE of Pt and the PMA in Co$/$Ni interfaces. We performed AHE measurements which allow us to display the fully experimental phase diagram of the magnetization switching. Using MOKE we imaged the magnetization reversal. The basic DW configuration is the Néel DW due to the DMI at Co$/$Pt interface which are nucleated at the edge of the bar due to the {\O}rsted field. We discussed the role of the in-plane magnetic field needed to induce the complete reversal of the magnetization.
The critical current density to reverse the magnetization, 0.3 TA/m$^2$, is of the same order of the critical current density for magnetization reversal in planar or perpendicular magnetic tunnel junctions \cite{Prejbeanu2013}, and lower than the critical current density for magnetization reversal in Co$/$Ni based metallic pillars \cite{Mangin2006a} and stripes \cite{KoyamaT.2012}, showing a better efficiency for potentials applications. There is still a room for theory and experiments towards a better understanding of the magnetization reversal and the way to reduce the necessary current density, which is the ultimate goal for technological applications.
\begin{acknowledgments}
This Work was partly supported
by the French Agence Nationale de la Recherche (ANR) through project
SOSPIN (2013-2016). We thank Cyrile Deranlot for fruitful discussions.
\end{acknowledgments}
\bibliographystyle{aipnum4-1}

\end{document}